\documentclass[
    ,final            
  ]
  {aipproc}

\layoutstyle{8x11single}

\usepackage{SIunits}
\begin{document}

\title{Acrylic purification and coatings}

\author{Marcin Ku\'zniak$^*$ for the DEAP collaboration}{
  address={$^*$Queen's University, Department of Physics, Kingston ON, K7L 3N6 Canada}
}



\begin{abstract}
Radon (Rn) and its decay daughters are a well-known source of background in
direct WIMP detection experiments, as either a Rn decay daughter or an alpha
particle emitted from a thin inner surface layer of a detector could produce a
WIMP-like signal.
Different surface treatment and cleaning techniques have been employed in the past
to remove this type of  contamination. A new method of dealing with the problem has
been proposed and used for a prototype acrylic DEAP-1 detector. 
Inner surfaces of the detector were coated with a layer of ultra pure acrylic, meant
to shield the active volume from alphas and recoiling nuclei.
An acrylic purification technique and two coating techniques are described: a solvent-borne (tested on DEAP-1) 
and solvent-less (being developed for the full scale DEAP-3600 detector). 
\end{abstract}

\maketitle


\section{Introduction}
DEAP-3600 is a dark matter detector, currently under construction at SNOLAB, 
designed for a three year background-free run with a 1000 kg sensitive liquid argon target (single phase), 
with sensitivity to spin-independent WIMP-nucleon scattering with cross-sections as low as 10$^{-46}$ cm$^2$ per nucleon. 
DEAP-1, a 7 kg liquid argon prototype detector, has been operated underground for background studies since 2008.
Backgrounds coming from natural sources of radiation (neutrons, $\gamma/\beta$ and $\alpha$-particles) are 
the largest challenge of the project and are more generally discussed elsewhere~\cite{Bei}.
Both DEAP detectors essentially consist of an acrylic vessel (AV), coated from inside with a wavelength shifting film 
(tetraphenyl butadiene, TPB) and filled with LAr. The scintillation light from LAr, shifted into the visible range
is detected in external photomultiplier tubes.

One particular source of backgrounds is caused by alpha decays of radon decay daughters in a thin surface layer
of the acrylic vessel (see~\cite{Bei}).
As Rn diffusion length in acrylic is 0.11~mm~\cite{Wojcik}, it is planned to use a resurfacing robot 
to remove up to 1~mm of material from the inside of the AV. However, the material may become saturated with 
Rn even before the acrylic is cast: either during storage (as acrylic beads of small diameter) or as liquid monomer, 
before the actual polymerization. In that case, the background from residual bulk contamination would still be visible. 
Detailed knowledge of the history of particular stock and details of the industrial polymerization process are necessary 
to exclude the possibility of the bulk contamination.

A stringent requirement on the $^{210}$\/Pb content ($<10^{-20}$~g/g) in acrylic motivated us to: 1) put in place a quality assurance 
program with the acrylic supplier, 2) work on a sufficiently sensitive direct assay technique for $^{210}$Pb and 3) find a way to coat 
inner acrylic surfaces with a layer of ultra pure material, sufficiently thick to stop alpha particles coming from the decays in the bulk,
which is described in detail in this document. 

The ultra pure coating will be needed for DEAP-3600, if the acrylic purity or assay 
sensitivity prove insufficient to meet the requirements.
A successful method has been found and tested with DEAP-1, work is ongoing to scale it up for DEAP-3600.

\section{Requirements for the coating}\label{sec-requirements}
In order to find an optimal candidate for the coating material, the following list of requirements was taken into account:
\begin{description}
\item[Impenetrable for $\alpha$] The most likely background event source would be the 5.3~MeV alpha from $^{210}$\/Po decay, which has a typical range of 
several tens of microns, depending on material density and composition, e.g. 34~microns in acrylic~\cite{SRIM}. To set the scale, the highest alpha energy
available from U/Th decay chains 8.78~MeV alpha from $^{212}$\/Po decay (extremely unlikely to be present in the bulk), in acrylic has a range of 75~microns.
Thus depending on the coating material, a thickness ranging from 50 to 100 microns is necessary. 
\item[Transparency] High attenuation length for wavelength from the TPB emission spectrum (380--500~nm), refractive index closely matched to acrylic, i.e. around 1.5.
\item[Stability] The coating has to survive the cool-down to liquid argon temperature without delamination or any other damage, which translates to: coefficient of 
thermal expansion (CTE) closely matched to acrylic, also good adhesion to acrylic.
\item[Possible to purify] As distillation is the most effective purification technique, it is preferable to start the 
process with liquid compounds.
\item[Compatibility] Coating process and chemicals used should be sufficiently compatible with acrylic and should not excessively dissolve or induce crazing in the
substrate. Stressed PMMA is particularly sensitive to chemical attack and crazes when in contact with most common organic solvents.
\item[Non-scintillating] Not to introduce a new source of backgrounds.
\end{description}
The coating has to be applied between the TPB and the acrylic substrate. Overcoating the TPB is not feasible as the only
group of materials capable of transmitting VUV light (LAr scintillation peak is around 128~nm) are fluorides, such as MgF$_2$. 
The thickest stable coatings of that type are $<$1~micron thick~\cite{MgF2}, which is insufficient to shield alphas.

Requirement of mechanical stability and CTE matched to acrylic removes pretty much all other inorganic compounds, limiting the choice to other plastics/polymers
only. A conclusion was reached that the best coating material would be either PMMA itself or some other polymer from the same family as PMMA, 
i.e. some other poly methacrylate or poly acrylate. Most of these compounds closely match the acrylic index of 
refraction and have very similar CTE. The polymerization process is relatively simple: based on free radical polymerization of a liquid monomer, 
which can be either thermally or UV induced. Simple (meth-)acrylates do not contain aromatic rings and thus do not scintillate. Finally, the solubility of different 
polymers from the family in common solvents and also their blend miscibility with PMMA varies significantly\cite{Harton05}. It should be possible to find a 
combination of monomers (and possibly solvents) sufficiently gentle for an acrylic substrate, yet capable of depositing a stable layer of clean material.

\section{Coating Methods}
For polymethacrylate coatings several production methods are possible:
\begin{enumerate}
\item Solvent-borne. A solution of a polymer in an organic solvent is directly applied to the surface and 
left for drying. Application methods include: casting, brush painting, spraying, dipping 
and spin coating. 
\item Chemical vapor deposition. Developed fairly recently~\cite{ChanPhD, Chan05, Anthamatten08}. Monomer and 
initiator vapors are introduced to an oxygen-free reaction chamber with the substrate. The initiator is activated 
either thermally or optically and turns into free radicals, which sustain growth of a polymeric coating on the 
substrate surface. 
\item Hybrid approach (in-situ polymerization). The monomer with small admixture of a photoinitiator directly 
applied to the surface (as in the first method), polymerization is then induced thermally or optically.  
\end{enumerate}

For the DEAP-1 chamber, the solvent-borne option was chosen as technically the simplest one. Because of the 
cylindrical geometry, limited amount of material and required uniformity of the layer, spin coating was selected as the application method.
An important factor was also the
existing experience in production of $\sim$3~micron PS coatings with that technique~\cite[Sec.~5.4.2]{Kuzniak08} and reported
successful production of much thicker $\sim$100~micron PMMA coatings with rotation speed increased to 1000~RPM~\cite{Gao06}.
For the windows, which are flat acrylic discs, casting was selected as the easiest and potentially cleanest of all possibilities.

In case of DEAP-3600, solvent-borne methods are not recommended, as introducing large amount of solvent into the AV would
increase the risk of crazing (achievable PMMA concentrations are around 5--10\% only). This is not an issue for DEAP-1, where
acrylic is essentially stress free. Ongoing R\&D on scaling-up one of two remaining solvent-less methods 
is described later in the paper.

\section{Purification scheme}\label{sec-purification}
Water extraction, filtering, adsorption, distillation are well known ways to purify liquids. 
Generally, distillation can be considered the 
most effective one, which is a conclusion reached by many other low background experiments, including Borexino, KamLAND and SNO+. Also, 
for even better purity, distillation can be done multiple times or combined with adsorption on a porous material. 
Water extraction would not be feasible, as both most acrylic monomers and relevant solvents mix very well with water. 

Relevant contaminants in our case are $^{210}$Po and $^{210}$Pb, as we already know that U/Th content near our target 
specification should be feasible even with commercial acrylic. 

In terms of achievable purification factors, there is a reported factor of 10$^4$ reduction for 
$^{212}$Pb in linear alkyl benzene (LAB, T$_B$=282--302\celsius), achieved by vacuum distillation 
combined with adsorption on Al$_2$O$_3$ (alumina) 
powder~\cite[Sec.~4.5]{Quirk08,Lan07} and the reported reduction with the distillation alone is about 4 times lower. 
Also, three different adsorbants 
are compared (alumina, silica gel and HZrO loaded silica gel), with alumina being the best one~\cite[Sec.~4.4]{Quirk08}. 
The purification factor due to adsorption on alumina only was around 2000. 

More detailed studies on pseudocumene (PC, T$_B$=170\celsius) purification through adsorption on silica gel performed for 
Borexino and KamLAND indicate $>8.9$ reduction for $^{210}$Pb and $>380$ for $^{210}$Po in a small scale setup~\cite[Tab.~4.26]{Niedermeier05}. 
Reduction factors from the range of 10--100 either on silica gel or alumina were reported for $^{212}$Pb~\cite[Tab.~6.5]{Keefer09}.
Distillation yielded reduction factors up to 3$\times 10^4$ for $^{212}$Pb~\cite[Sec.~6.3]{Keefer09}, with the final content of around $10^{-25}$~g/g.

Data on Po removal through distillation is sparse and probably the best clue available is a purification 
factor of ``greater than 500$\pm$90 for $^{210}$Po removal'', 
reported in Ref.~\cite[Sec.~4.11]{Leung06}. Polonium is known to be more difficult to remove than Pb because of its non-trivial
volatility~\cite{Mabuchi58,Mabuchi63,Martin69,Hussain95,MurrayMatthews07,Momoshima01}. 
Some results indicate than adsorption cannot be considered the primary purification technique, because 
it is not effective against organometallic Pb and Po compounds~\cite[Sec.~6.3.7]{Keefer09}. 
Still, the general consensus in the field is that the distillation is the most efficient method, and there are some indications that adsorption could 
additionally improve 
the purity.
Moreover, because of the fact that boiling points of ACN (82\celsius) and MMA (101\celsius) are significantly lower than 
for LAB or PC, even greater purification factor should be achievable for DEAP.  
 
Purification of solids via distillation is a little more problematic as it requires
both highly reduced pressure and high temperature. No published data on purification efficiency of the wavelength shifter was found up to date. 
However, the vacuum evaporation process used to deposit TPB
coatings in DEAP is effectively equivalent to distillation, so should result in a substantial purification of the wavelength shifter.
Therefore no dedicated distillation setup was planned, although certain improvements described later were added
to the evaporation procedure. A dedicated TPB distillation setup is certainly planned for the future application in DEAP-3600.

\section{PMMA coating for DEAP-1}\label{sec-selection}
\paragraph{Initial tests}
Achievable concentrations, drying time and cast coating quality and uniformity were investigated for several common solvents: acetone, acetylonitrile 
(ACN), tetrahydrofuran (THF), toluene and, finally, the methyl methacrylate monomer (MMA). ACN and THF turned out to be the best, allowing to use 
concentrations close to 10\%. Eventually ACN was chosen, as THF might introduce some additional safety hazards during 
distillation (due to its tendency to form explosive peroxides). It was also observed that solubility strongly depends on details 
of polymerization process (high level of cross-linking reduces the solubility).

\begin{figure}[htb]
\includegraphics[scale=0.7]{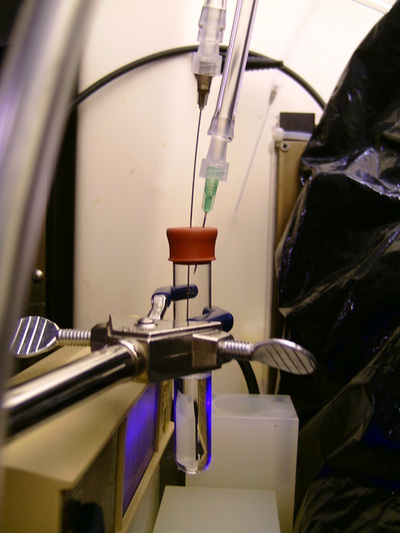}
\caption{UV induced polymerization in a test tube constantly purged with N$_2$.\label{fig-test-pol}}
\end{figure}

Commercially available MMA contains small amount of inhibitor (typically hydroquinone), which reduces the risk 
of spontaneous, egzothermic polymerization of large quantities of MMA. The inhibitor is removed via dripping the monomer through 
a filtration column filled with alumina powder (Al$_2$O$_3$). Also oxygen is a very potent polymerization inhibitor, 
thus the polymerization process has to be performed in an inert N$_2$ atmosphere. Figure~\ref{fig-test-pol}
shows a typical polymerization test setup. 

Polymerization has to be started with free radicals, typically produced by break-up of initiator molecules. Although MMA 
and some other methacrylates can self-polymerize when illuminated with short wave UV radiation (around 254~nm), i.e. act as their own initiators, 
the resulting polymer is highly cross-linked and, in consequence, insoluble.
Initiators are usually very reactive, sometimes unstable and tend to decompose at elevated temperatures, thus are not suitable for 
distillation. Majority of common initiators are solids at room temperature, with a notable exception of azoinitiators, 
such as easily available 2,2'-azobis(2-methylpropane), ABMP (also called azo-tert-butane), our final choice.
Required initiator concentration is usually of the order of 0.1\%, therefore its purification via adsorption would overall provide a sufficient 
purification factor (see Sec.~\ref{sec-purification}). Another advantage of azoinitiators is that their decomposition into free radicals is induced
with long-wave UV light (366~nm), which is not damaging to the base acrylic~\cite{Chan05}.  

The spin coater was designed to operate at 500 RPM rotational speed with the DEAP-1 acrylic insert for several hours.
Thin teflon foil rings were used as a soft seal between Delrin endcaps (attached to a rotating rod) 
and the insert. 
\begin{figure}[htb]
\includegraphics[scale=0.4]{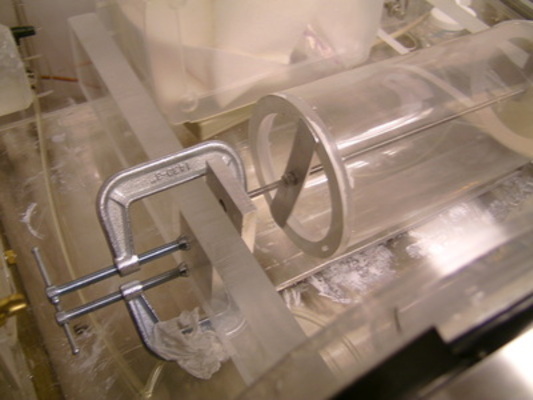}
\includegraphics[scale=0.4]{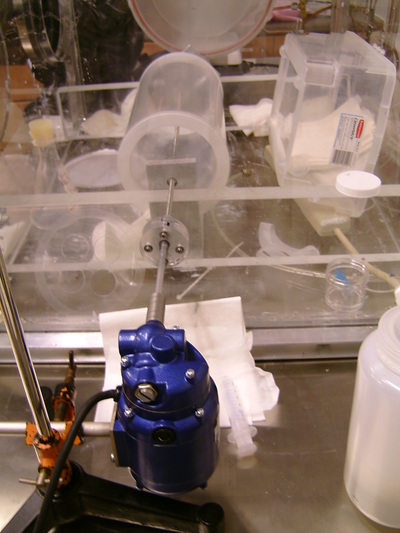}
\caption{Spin coater. On the left: installed in the glovebox, frame of the spin coater attached to acrylic beams with 
stainless steel clamps. On the right, outside of the glovebox: the motor and the mechanical feedthrough.\label{fig-spincoater}}
\end{figure}
The spin coater was installed in a glovebox, held rigidly by two acrylic bars. 
The motor was attached to the table outside of the glovebox, with a feedthrough for the rotating rod installed in the wall 
of the glovebox. An external motor had to be used, in order to minimize Rn emanation inside the glovebox. All parts of the 
spin coater were ultrasonically washed before installation in the glovebox. The complete assembly is shown in 
Fig.~\ref{fig-spincoater}.

\paragraph{Final production}\label{sec-production}
In order to obtain $\sim$100~\micro\metre\ thick coating for inner DEAP-1 surfaces, including some reasonable safety margin
about 23~g of purified acrylic were necessary. In order to have some extra material for witness samples, it was planned to 
distill about 30~ml MMA and 450~ml of ACN.
\begin{enumerate}
\item Solvent distillation.
2/3 of the large distillation flask (1~l capacity) of HPLC grade ACN was used for the distillation with addition of 
Al$_2$O$_3$. The distillation was performed at $\sim$280~Torr pressure and around 50\celsius\ temperature 
(with constant N$_2$ purge, see Fig.~\ref{fig-test-dist}).
\begin{figure}[htb]
\includegraphics[scale=0.7]{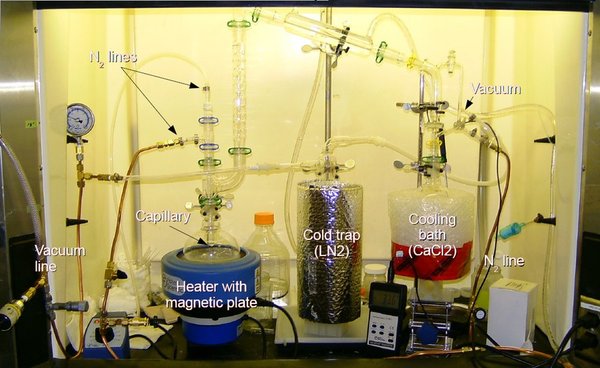}
\caption{The distillation apparatus at Queen's.\label{fig-test-dist}}
\end{figure}
\item Monomer distillation.
MMA taken for the distillation had already been dripped through a filtration column with alumina, in order to remove the inhibitor. 
During the distillation, as an additional precaution to minimize risk of spontaneous polymerization inside the apparatus, its 
'hot' parts and the condenser were covered with black plastic foil and the fume hood lamp was not used. 
The distillation was performed at $\sim$160~Torr pressure and around 50\celsius.
\item Polymerization.
Small amount of alumina and a small magnetic stirrer was added to a vial containing ABMP, and the mixture was stirred
for several hours inside the glove box. Then it was opened, sucked into a clean syringe and pushed through 0.2~\micro\metre\ PTFE
filter into the flask with distilled monomer. After re-sealing (with a tap), the flask was taken out of the glove box, attached 
to a N$_2$ purge line and illuminated with an UV lamp (PASCO OS-9286A). Gas purge was used during the polymerization.
After only 30 minutes, the material was already fully polymerized. The flask was resealed and taken back to the glove box.
Chunk of PMMA was then fragmented into smaller pieces with a pair of clean stainless steel nippers, 
put in a 500~ml cleaned and etched Erlenmeyer flask with distilled ACN and a teflon magnetic stirrer. The mixing was speeded up with a NiCr heater. 
\item Applying.
Acrylic windows were levelled and fixed inside large diameter band clamps sealed with PTFE tape before introducing them
to the glove box. Dissolved acrylic was poured on their surfaces from a graduated cylinder, distributed uniformly and left 
for drying. 
For spin-coating about a half of acrylic paint was poured into the acrylic insert, which was then slowly rotated several times in 
order to wet the entire inner surface, then the rotation speed was slowly ramped up above 500~RPM. 
After a total time of 5h30m the coating was transparent, 
uniform and dry (see Fig.~\ref{fig-final}). Based on the amount of paint used, the approximate estimate of the coating thickness is 80~\micro\metre. 
\begin{figure}[htb]
\includegraphics[scale=0.4]{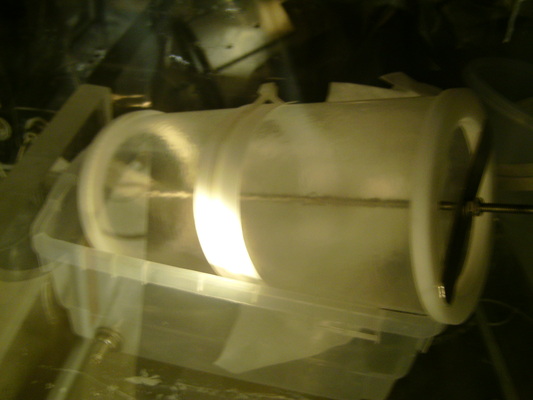}
\caption{DEAP-1 acrylic insert with a solidified coating, after disassembling the spin coater. \label{fig-final}}
\end{figure}
Windows and the insert were then outgassed in vacuum for several days.  
\end{enumerate}

\section{Scale-up for DEAP-3600}
The inner surface area of DEAP-3600 is approximately 9~\metre\squared. A 100~\micro\metre\ thick acrylic coating on that
surface translates into about 1~kg total mass of the extra layer. Regardless of technical details of the application,
the main difficulty with solvent-borne methods is the necessity to evaporate $>$10~liters of solvent from the acrylic vessel (AV), 
which is a very confined volume. In consequence, the acrylic vessel would have to be exposed to aggressive solvents for an extended period 
of time, which could increase the risk of crazing. This motivated us to search for an alternative method and test it in a simplified geometry. 
In both cases described below, the plan is to eventually integrate coating applicators as with the resurfacer robot, planned for DEAP-3600.
\paragraph{Chemical Vapor Deposition}
Chemical vapor deposition (CVD) occurs when a coating is produced as an effect of reactions occuring in gas phase (in a 
controlled mixture of a precursor gas and an initiator). 
The method has been known for a long time and typically used to deposit inorganic layers of silicon, oxides, nitrides or different types
of carbon structures, including diamond. Usually the process requires very high temperatures (well beyond 500\celsius).

At the beginning of the decade, which is fairly recently, a group from MIT has tuned and optimized the hot filament CVD (HFCVD)
method for production of thick PTFE and other polymeric coatings~\cite{Lau03,PryceLewis09}. Deposition rates of 
$\sim$1~\micro\metre\per min have been reported and the method had a good potential for scalability.

A detailed experimental and theoretical study was later done on HFCVD of poly methacrylate polymers~\cite{ChanPhD}, including
PMMA. Also, instead of hot filaments, a combination of UV light and UV sensitive initiator (ABMP) was also successfully 
tested~\cite{Chan05}. For poly methacrylates, generally, deposition rates up to 100~nm\per min were reported in those 
pioneer experiments, proportional to the UV light intensity. Since the deposition rates were anti-correlated with the monomer saturated vapor pressure,
heavier monomers, such as e.g. cyclohexyl methacrylate (CHMA) would result in significantly higher deposition rates than methyl methacrylate (MMA).

Further study was performed in a modified geometry~\cite{Anthamatten08}, with a cylindrical deposition head instead of
a matrix of hot filaments.

Given all that information, it seems feasible to develop a system capable of overcoating the AV on the time scale of up to 
several weeks. In terms of the deposition rate, CHMA was chosen as the most promising monomer, also the UV induced CVD 
(because of the potential of increasing the rate with higher light intensity). Operating at the atmospheric pressure 
is desirable for the resurfacer vacuum compatibility reasons.

\begin{figure}[htb]
\includegraphics[scale=0.4]{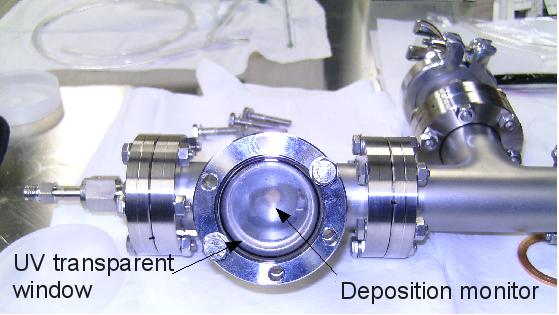}
\caption{Test setup for photoinitiated CVD. \label{fig-cvd}}
\end{figure}

\paragraph{CVD tests at Queen's}
A small test CVD chamber was constructed at Queen's at the end of 2009 in order to achieve the deposition rate of at least 
1~\micro\metre\per\minute\footnote{A requirement based on two 100~\centi\metre\squared\ coating heads operating inside AV
for several weeks}. It consists of a couple of CF tees, with a viewport 
(glass window or UV transparent FEP foil) and a water cooled deposition monitor located 2 inches below (see Fig.~\ref{fig-cvd}). 
Two N$_2$-purged bubblers, one with the monomer and the other with the initiator, are connected to the chamber and 
there is also a water filled outlet bubbler to prevent ambient air (especially oxygen) from entering the volume.

\begin{figure}[htb]
\includegraphics[scale=0.25]{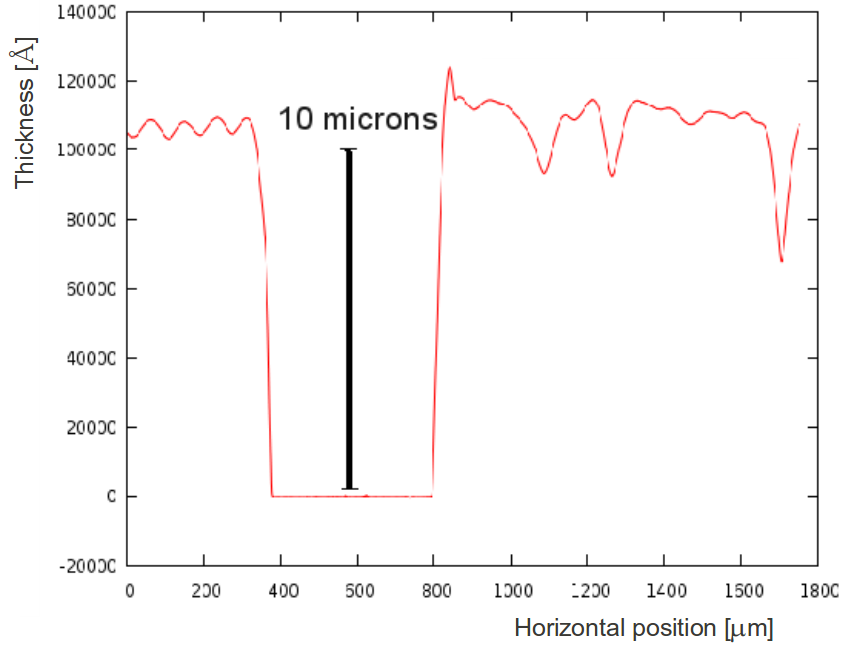}
\includegraphics[scale=0.25]{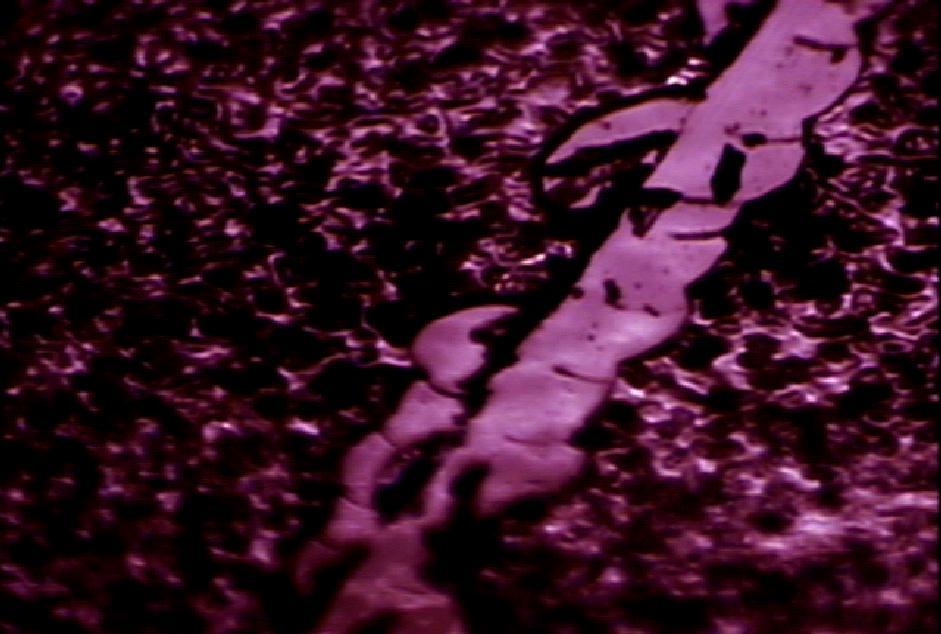}
\caption{On the left a profiler scan across a scratch made in the coating with a scalpel. Flat surface at the bottom
of the scratch belongs to the glass substrate. On the right, the 
microscope picture of the scanned area.\label{fig-pchma}}
\end{figure}

After the initial purge, the monomer bubbler temperature would be ramped up to $\sim$80\celsius\ and the light
source would be turned on. Effect of different flow rates through both bubblers on the deposition rates
could be observed thanks to the deposition monitors. In some cases a couple of additional glass slides were
introduced to the chamber around the deposition monitor in order to measure the coating thickness directly with a 
profiler. Figure~\ref{fig-pchma} shows a typical scan and a microscope picture of one of the first coatings deposited.
A maximum deposition rate of about 100~nm/min was achieved at that time, with a PASCO OS-9286A light source.

Since that time several changes were introduced to the setup. Most importantly, the mercury UV light source was changed 
to an UV LED head\footnote{OmniCure LX400 spot curing system from EXFO.}, capable
of delivering $\sim$1~W\per\centi\metre\squared\ at the substrate surface. The new source not only provides much more
intensity, its bandwidth (365$\pm$20~nm) is better tuned with the characteristics of ABMP (the UV initiator). Additionally,
to maximize the intensity even further, the glass window in the viewport was replaced with a UV transparent FEP foil.
Varying the temperature of the bubbler with the initiator was also tested (between -20\celsius\ and the room temperature).
It was also tried to deliver the initiator vapor very close to the surface, to prevent earlier mixing with the monomer vapor 
(a thin teflon tube inside the chamber was added).

Tests are still ongoing, but some preliminary conclusions can be already made. Deposition rates slightly above 
1~\micro\metre\per\minute\ were achieved, however, a larger increase was expected, assuming that the UV intensity 
was the only scaling factor. At that rate other limiting factors may come into play, such as the rate at which 
the monomer vapor is delivered to the chamber and is then adsorbed on the substrate surface.
If further tests confirm this, the conclusion can be made that the CVD method would work for DEAP-3600, although at the limit
of its feasibility. It also means that another, possibly even easier approach could be more effective (see below).

\paragraph{Outlook}
Since the material transport rate might be the bottle neck of the process, it was recently proposed to test a hybrid method:
spray the monomer/initiator mixture on the inner AV surface and then UV polymerize it. Tests performed on highly stressed acrylic 
dogbones indicate that crazing in contact with CHMA starts after several hours, to be compared with seconds in case of MMA.
Therefore spraying a thin layer of liquid on the AV surface and curing it within 30--60 minutes should be sufficiently safe. With
no solvent used in the process, the total amount of liquid to deal with is on the order of one liter for the entire AV.

An appropriate supplier and a spraying system capable of depositing very thin and uniform layers have been identified, 
it is an airless automated industrial gun, with both the monomer and compressed gas (for valve control) delivered at 30 
psi pressure. Its specification make it easy for incorporation in the resurfacer. 

\section{Summary}
The risk of an additional source of background caused by bulk contamination of acrylic with radon daughters and 
resulting alphas has been identified. 
The ultra pure coating may be necessary for the experiment if the acrylic assay results are not satisfactory.

The proposed purification method is based on distillation and adsorption and has been widely used in low background 
physics experiments for other chemicals. A purification setup has been constructed at Queen's, successfully used and
is available for the future (it would be sufficient for the full scale purification).

A solvent-based spin coating method has been developed and used to produce a new DEAP-1 chamber. 
Technically, the method worked very well, although no significant improvement was reached in DEAP-1 background rate. 
Evidence was found that currently a dominant fraction of the rate is still induced by radon and thoron emanation in 
the process systems. Other methods are being developed to remove these sources and the level of the bulk contamination 
in acrylic remains an open question.

Chemical vapor deposition has been considered an interesting option for DEAP-3600. It was managed to produce multi-micron
thick coatings using a small CVD chamber and achieve a high deposition rate. Tests are still ongoing. There is also
a new, possibly more robust technique, which we are starting to work with. It is based on spraying the liquid monomer on 
the surface and then UV curing it. Last series of optimization steps will be possible with a new test setup available shortly.
The final design of the applicator, based on test results, is expected in the first half of 2011.



\begin{theacknowledgments}
I would like to thank everyone who contributed to this work, especially: 
Chris Jillings and the SNOLAB team (DEAP-1 deployment and operation), Tina Pollmann (TPB evaporation),
David Bearse, Garry Contant, Robert Gagnon, Charles Hearns, Bernard Ziomkiewicz (for technical support),
and Mitchel Anthamatten, Robin Hutchinson, Tomasz Moszczy\'nski and Kevin Robbie 
for discussions and advice.
This work is supported by the National Science and Engineering Research Council of Canada (NSERC), by the Canada
Foundation for Innovation (CFI), by the Ontario Ministry of Research and Innovation (MRI), by Queen's University 
and by the David and Lucille Packard Foundation.
\end{theacknowledgments}


%
\bibliographystyle{aipproc}   





\end{document}